%% file: elsarticle-template-num.tex
\documentclass[fleqn,final,5p,times]{elsarticle}
\usepackage{amssymb}
\usepackage{calc}
\usepackage{siunitx}
\usepackage{physics}
\usepackage[hidelinks]{hyperref}
\usepackage{cleveref}
\usepackage{xcolor}
\usepackage{xargs}
\usepackage{soul}
\usepackage{subcaption}
\usepackage{macros}
\usepackage{booktabs}
\usepackage{tabularx}
\usepackage{mathtools}
\DeclareMathAlphabet{\mathpzc}{OT1}{pzc}{m}{it}
\usepackage{algorithm}
\usepackage{algpseudocode}

\usepackage[mode=buildnew,subpreambles=true]{standalone}
\usepackage{shellesc}
\usepackage{tikz,tikzscale}
\usetikzlibrary{arrows,calc,fit,patterns,positioning,spy,backgrounds,decorations.fractals,shapes.misc}
\usetikzlibrary{shapes.geometric}
\tikzset{boximg/.style={remember picture,black,thick,draw,inner sep=0pt,outer sep=0pt}}
\usepackage{pgf}
\usepackage{pgfplots}
\usepgfplotslibrary{statistics}
\usepgfplotslibrary{fillbetween}
\usepgfplotslibrary{colormaps}
\pgfplotsset{compat=1.18}
\pgfplotsset{lua backend=true}

\pgfplotsset{select coords between index/.style 2 args={
    x filter/.code={
        \ifnum\coordindex<#1\fi
        \ifnum\coordindex>#2\fi
    }
}}

\usepackage{graphicx}
\usepackage{xspace}
\newcommand{\hawk}{\textit{Hawk}\xspace}

\newcommand{\elexi}{\reflectbox{E}LEXI\xspace}

\journal{International Journal of \LaTeX templates}

\allowdisplaybreaks

\begin{document}
\sloppy

\begin{frontmatter}

\title{High-Order Simulation of Particle-Laden Flows in Moving Domains Using Coupled ALE and Sliding Mesh Approaches}

\author[iag]{Anna Schwarz\corref{cor}\fnref{fn}}
\ead{schwarz@iag.uni-stuttgart.de}
\author[iag]{Patrick Kopper\corref{cor}\fnref{fn}}
\ead{kopper@iag.uni-stuttgart.de}

\cortext[cor]{Corresponding author}
\fntext[fn]{P. Kopper and A. Schwarz share first authorship.}

\affiliation[iag]{organization={University of Stuttgart, Institute of Aerodynamics and Gas Dynamics},
            addressline={Wankelstr. 3},
            city={Stuttgart},
            postcode={70563},
            country={Germany}}

\begin{abstract}
In practical applications, compressible particle-laden flows in moving geometries involve complex, non-linear, and multi-scale interactions with turbulent structures. Resolving these dynamics numerically requires careful algorithmic treatment to accurately predict particle trajectories. This work presents a high-fidelity Euler--Lagrange framework that couples a high-order discontinuous Galerkin spectral element method for the continuous phase with a Lagrangian point-particle tracking scheme. To manage moving and deforming domains, the framework integrates two distinct mesh movement strategies: the arbitrary Lagrangian--Eulerian method for general mesh deformations such as time-resolved particle-induced surface deformations and its special interface case, the sliding mesh approach, uniquely suited for rigid rotational or translational movements. A primary focus is placed on tightly coupling the arbitrary Lagrangian--Eulerian formulation into the temporal evolution step by utilizing radial basis function morphing to capture the non-linear feedback loop between evolving surface topologies and the continuous phase. Concurrently, the framework ensures time- and high-order accurate coupling of the mesh movement algorithms with the dispersed phase. In particular, the proposed algorithm resolves the sliding mesh tracking problem by enforcing strict spatial and temporal accuracy as Lagrangian particles cross non-conforming grid interfaces between adjacent moving zones. The algorithms are rigorously validated against multiple benchmarks and subsequently applied to two challenging compressor rotor applications: the first focusing on solid-particle erosion, and the second featuring an upstream cylindrical wake generator.
\end{abstract}


\end{frontmatter}

\input{1_introduction}

\input{2_theory}

\input{3_numerical_methods}

\input{5_validation}

\input{7_application}

\input{8_conclusion}

\section*{Acknowledgements}
The research presented in this paper was funded in parts by Deutsche Forschungsgemeinschaft (DFG, German Research
Foundation) under Germany's Excellence Strategy - EXC 2075 - 390740016 and by the European Union.
This work has received funding from the European High Performance Computing Joint Undertaking (JU) and Sweden, Germany, Spain, Greece, and Denmark under grant
agreement No 101093393.
We acknowledge the support by the Stuttgart Center for Simulation Science (SimTech).
The authors gratefully acknowledge the support and the computing time on \hawk provided by the HLRS through the project
``hpcdg''.

\bibliographystyle{elsarticle-num-names}
\bibliography{references}

\end{document}

%% file: 1_introduction.tex
\section{Introduction}%
\label{sec:introduction}

High-fidelity simulations of particle-laden turbulent flows in moving domains, such as those found in turbomachinery applications~\cite{Ghenaiet2010}, must handle the complex coupling between multi-scale, high-velocity fluid dynamics and transient particle trajectories.
To accurately capture these localized physics and understand their impact on macroscopic phenomena like surface erosion, component
degradation, and phase distribution, predictive numerical modeling has become an indispensable tool~\cite{Beck2019a,Hartmann2022}.
These challenges have driven the development of numerous modeling approaches aimed at enabling high-fidelity simulations of such particle-laden flows.
One prominent approach is the Euler--Lagrange or point-particle method, in which the continuous phase is resolved in an Eulerian
frame of reference while the particles are tracked as discrete points evolving in a Lagrangian manner according to Newton's second
law of motion. Compared to Euler--Euler frameworks, Euler--Lagrange methods are able to manage large counts of disparate particles with high
accuracy while circumventing the prohibitive computational costs associated with fully particle-resolved strategies~\cite{Balachandar2010}.
When coupled with a high-order approach such as the discontinuous Galerkin (DG) method, Euler--Lagrange methods provide high-fidelity solutions for turbulent flows with millions of particles~\cite{Ching2021,Schwarz2025a}.

However, the classical point-particle formulation assumes a static spatial domain; when boundaries move, the primary computational challenge shifts to accurately tracking these particles across dynamically altering mesh topologies.
Depending on the problem under consideration, different numerical methods are best suited to handle the underlying grid deformation or motion.
In this paper, the primary focus is on the integration of particle tracking within two of the most prominent mesh movement algorithms:
the arbitrary Lagrangian--Eulerian (ALE) method~\cite{Minoli2011,Wang2015,Schwarz2025b} and a special case of ALE, the sliding mesh approach~\cite{Murphy1994,Duerrwaechter2021}. While Chimera (overset) grid techniques~\cite{Wang2014} are also commonly applied to handle complex multi-body movements in particle-laden flows~\cite{Aarnes2019}, they are not utilized in the present work.

The ALE method, which tracks moving boundaries dynamically through displacement of the mesh nodes, is highly effective for localized or arbitrary mesh deformations, which are permitted to be arbitrary under the condition that mesh quality is maintained.
To prevent highly skewed or unphysically collapsed mesh elements, this framework is typically coupled with volume-mesh morphing techniques
such as a radial basis function (RBF) interpolation~\cite{DeBoer2007} or inverse distance weighting to smoothly propagate surface
boundary deformations into the interior volume. Transformation of the continuous phase from the physical, moving domain to a static
reference domain, as commonly done in DG methods, introduces a metric evolution equation, known as the geometric conservation law (GCL)~\cite{Lesoinne1996}. This condition ensures that the continuous and discrete conservative properties of the governing equations are perfectly preserved. While the ALE procedure inherently captures complex macroscopic variations like fluid-structure interactions~\cite{Krais2020} or large surface deformations that exceed statistical roughness models~\cite{Lo2023}, it requires the continuous recomputation of grid metric terms and is generally limited to moderate deformations. Consequently, for the discrete phase, this continuous spatial displacement of mesh nodes demands a robust localization algorithm capable of accurately tracking particle positions relative to the moving grid cells, particularly near boundaries.

Conversely, the sliding mesh approach is utilized for domains undergoing large-scale, rigid translational or rotational movements, where
distinct mesh zones slide past one another along a well-defined non-conformal interface~\cite{Duerrwaechter2021}. For simplification, this paper does not consider arbitrary mesh deformation within the subdomains in the context of the sliding mesh approach, thus the individual grid cells do not deform. As such, the sliding mesh approach will be considered a special case of the ALE method for a-priori known translational or rotational rigid body movements without the need to recompute metric terms. Instead, the connectivity across the interface is updated dynamically at every time step to maintain flux conservation for the continuous phase. For the discrete phase, this architecture requires a robust intersection algorithm to seamlessly transition Lagrangian particles across the sliding interface from one moving frame to another. This non-conformal crossing introduces sharp spatial discontinuities in the particle trajectory, rendering the algorithmic handling of the interface a critical bottleneck for particle tracking accuracy.

Consequently, integrating a Lagrangian particle tracking scheme into either of these moving grid approaches demands exact temporal and spatial interpolation at the moving interfaces.
Particle trajectories are subject to intense temporal variations induced by both turbulent flow structures and the unsteady interactions associated with moving boundaries.
High-fidelity simulations that combine these mesh-motion techniques with advanced Euler--Lagrange frameworks are therefore essential for
resolving transient accumulation, impact, and back-scattering phenomena that cannot be captured by simplified steady-state approaches.
While segregated approaches for dynamic grid movement and particle-surface interactions are well-established, literature detailing
high-order Euler–Lagrange tracking on simultaneously moving, deforming, and non-conforming curvilinear meshes is, to the authors'
knowledge, exceptionally limited. Recent efforts, such as the parallel algorithm by~\cite{Xiao2023} or the multi-zone approach by~\cite{Morelli2019} for tracking particles across sliding non-conformal interfaces and/or deforming rigidly moving meshes, are restricted to a low-order finite volume (FV) discretization for the continuous phase. Furthermore, although particle-resolved methods utilize ALE or Chimera-type approaches to resolve the particle-fluid interface, these methods focus on the dispersed phase geometry and are not designed for the point-particle coupling within moving, non-conforming curvilinear meshes, e.g.~\cite{Maxon2021}.

Motivated by this gap in the literature, this paper aims to present a high-order consistent, time-accurate particle tracking
framework on moving and deforming, curvilinear meshes using the arbitrary Lagrangian--Eulerian method and a special case of ALE, the non-conformal sliding mesh interface. Special focus is placed on preserving conservation properties, high-order accurate particle tracking, and computational efficiency.
In this work, the ALE method is applied to arbitrary surface deformations, while the sliding mesh procedure handles translational or rotational rigid-body motion across non-conforming interfaces.
To maintain parallel scalability, the ALE framework interpolates surface displacements into the volume using compactly supported RBFs constructed locally on each compute core. Particles are tracked within an absolute frame of reference, which naturally incorporates the mesh velocity and ensures accurate trajectories without additional correction terms. For the sliding mesh framework, special care is taken to maintain high-order temporal and spatial accuracy as Lagrangian particles transition across the non-conforming interface.

The proposed algorithms are code-agnostic, facilitating straightforward integration into any existing Euler--Lagrange framework.
For this study, the methods are implemented within
\elexi\footnote{\url{https://github.com/flexi-framework/elexi}}~\cite{Kopper2023,Schwarz2025a}, a high-order Euler--Lagrange solver where the carrier phase is discretized using a hybrid operator that combines a high-order accurate discontinuous Galerkin spectral element method (DGSEM) with a localized, low-order finite volume operator.
The discrete phase is tracked simultaneously using a Lagrangian point-particle approach.
Consequently, the new algorithm seamlessly leverages \elexi's native capabilities for handling complex geometries on unstructured grids, including curved elements, hanging nodes, and advanced boundary conditions.
Combined, these features enable an efficient, high-fidelity treatment of the complex, coupled interactions between suspended particles, mesh movement, and the compressible carrier phase.

The primary focus of this work is the numerical formulation, implementation, and application of methods handling discrete particles within moving geometries under compressible flow conditions.
The remainder of this paper is structured as follows.
In~\cref{sec:theory}, the underlying equations for both the continuous and discrete phase are introduced, alongside the treatment of particle-wall interactions and particle-induced surface deformations.
A brief outline of the numerical treatment and algorithmic implementation of these equations is provided in~\cref{sec:methods}.
The framework is validated in~\cref{sec:validation}, followed by a demonstration of the solver's capabilities
in~\cref{sec:application} using two challenging, real-world engineering applications.
Finally,~\cref{sec:conclusion} offers concluding remarks.

%% file: 2_theory.tex
\section{Theory}%
\label{sec:theory}

\subsection{Continuous Phase}%
\label{sec:theory:nse}
The fluid field is governed by the compressible unsteady Navier--Stokes--Fourier equations.
Defined on the time-varying domain $\Omega_t \subset \R^3, \ t \in \R^+_0$, the equations are expressed in conservative form as
\begin{align}
  \frac{d\cons}{dt} + \nabla\cdot \fphys\left(\cons,\nabla\cons\right) = \mathbf{0},
  \label{eq:theory:fluid:NSE}
\end{align}
where $\cons = \smash{[\rho,\rho u_1,\rho u_2,\rho u_3,\rho e]^T}$ represents the vector of conservative variables with
$\rho$ as the fluid density, the $i$-th component of the velocity vector $u_i$ and the total energy per unit volume $e$.
The physical flux $\fphys$ accounts for both inviscid (Euler) and viscous contributions.
To facilitate the numerical treatment of the moving domains, \cref{eq:theory:fluid:NSE} is transformed to a time-invariant reference domain, $E = [-1,1]^3$ with $\cons(\refpos,\tau)$, by
introducing a differentiable mapping $\boldsymbol{\chi}: E \to \Omega_t, (\refpos,\tau) \mapsto (\pos,t)$, with time $\tau \in \R^+_0$.
Under this ALE formulation, the governing equations are reformulated as
\begin{align}
  \lr{\J \cons}_t + \gradientXI \cdot \fphysref  = \mathbf{0}, \hspace{0.5cm} \fphysref = \M^\transpose \lr{\fphys - \cons \otimes \meshVel},
  \label{eq:theory:nse}
\end{align}
where $\J=\mathrm{det} \Jm$ denotes the determinant of the Jacobian and $\M=\mathrm{adj} \Jm^\transpose \otimes \unit_{\nq}$ is the adjoint of the
Jacobian matrix $\Jm$ of the mapping.
The mesh velocity is defined as $\meshVel = \boldsymbol{\chi}_\tau \in \R^3$, and the geometric evolution of the elements is
constrained by the geometric conservation law (GCL), given as
\begin{align}
  \J_t &= \divXI \lre{\lr{\mathrm{adj}~\Jm}^\transpose \cdot \meshVel}.
  \label{eq:theory:gcl}
\end{align}
Here, $\tau = t$ (just a change of the notation) represents the time coordinate in the reference frame.
To ensure that the metric identities $\divXI \M = 0$ are satisfied at the discrete level, a prerequisite for maintaining
free-stream preservation, the metric terms $\M^\transpose$ are approximated using the conservative curl form \cite{Kopriva2006}.
Further details regarding the ALE derivation can be found in \cite{Minoli2011,Schnucke2018}.
The equation system is closed by the equation of state of a calorically perfect gas. The dynamic viscosity $\dynvisc$ is obtained from Sutherland's
law~\cite{Sutherland1893}, while the heat flux is modeled by Fourier's law. In accordance with Stokes' hypothesis, the bulk viscosity is set to zero.

\subsection{Dispersed Phase}%
\label{sec:theory:maxey}
In the Lagrangian point-particle framework, particles are modeled as discrete points characterized by mass $\partmass$ and virtual diameter $\partdiam$.
Their trajectory and physical state are governed by a system of ordinary differential equations (ODEs), written as
\begin{align}
  \frac{d \partpos}{dt} &= \partvel, \nonumber \\
  \partmass \frac{d \partvel}{dt} &= \mathbf{F}_D = 3\pi\dynvisc \partdiam \dragfactor \left(\fluidvel - \partvel \right). \label{eq:mrg:mom}
\end{align}
The particle position in physical space $\smash{\partpos=[\partpos[][1], \partpos[][2], \partpos[][3]]^T}$ and velocity $\partvel$
are determined by integrating the set of ODEs.
Following the Maxey–Riley–Gatignol (MRG) formulation \cite{Maxey1983,Gatignol1983}, the momentum equation \eqref{eq:mrg:mom} is simplified by considering only the drag force.
To account for high particle Reynolds numbers, an empirical correction is incorporated via the drag factor $\dragfactor$, which is computed according to \cite{Loth2021}.

\subsection{Intersection of Particles with Solid Walls}%
\label{sec:theory:rebound}
Particle-wall collisions are modeled via the hard-sphere approach, i.e., they are not resolved in time but handled in an a
posteriori manner. Thus, the timestep $\dtstage$ of the current Runge--Kutta (RK) stage has to be greater than the time a particle
requires to collide with a wall ($\dtstage > \dtcoll$). The change in particle momentum is given as
\begin{align}
  \partmass^{ref} (\partvel^{ref}  + 2 (\partvel^{imp} \cdot \normalvec) \normalvec)- \partmass^{imp} \partvel^{imp} = \partmomentum,
  \label{eq:theory:rebound}
\end{align}
where the superscripts $(\cdot)^{imp}$ and $(\cdot)^{ref}$ denote quantities immediately before and after the impact, respectively, and
$\normalvec$ represents the unit normal vector of the boundary in physical space.
While \cref{eq:theory:rebound} defines the conservation of momentum, the energy dissipation occurring during the impact is not inherently considered.
In physical systems, collisions are typically non-elastic, meaning a portion of the kinetic energy is lost.
In this context, the term $\partmomentum$ accounts for momentum changes due to these non-ideal effects. For the perfectly elastic
(reflective) collisions primarily considered here, $\partmomentum = \Null$, whereas for scenarios involving plastic
deformation, $\partmomentum \neq \Null$.

To close \cref{eq:theory:rebound} for $\partmomentum \neq \Null$ and in turn determine the post-impact states $\smash{x^{ref}}$, rebound models are employed which relate the pre- and post-collision kinematics.
The central parameter in these models is the coefficient of restitution (CoR), defined as the mapping $\cor{x} \colon x^{imp}
\mapsto x^{ref}$ from the impinging to the reflected value, given as
\begin{align}
  \cor{x}(\zeta) = \frac{x^{ref}}{x^{imp}}.
\end{align}
The coefficient $\cor{x}$ is generally a function of several parameters $\zeta$ that characterize the collision
physics and can be computed using adequate rebound models, here the rebound model of~\cite{Schwarz2022} is used.
The variable $x$ are usually the particle velocities in normal, $\cor{n}$, and tangential directions, $\cor{t}$.

\subsection{Particle-Induced Wall Deformations}%
\begin{figure}[!tb]
  \centering
  \includegraphics[width=0.7\columnwidth]{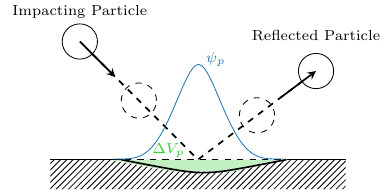}
  \caption{Sketch of the particle-induced wall-deformation using RBFs to model the distributed displacement induced by a single
  impacting particle around the impact point.}
  \label{fig:theory:rbf_ale}
\end{figure}

\label{sec:theory:partdeform}
Each particle impact modifies the local geometry of the target surface, either by removing material (erosion) or adding it (deposition), depending on the impact energy and material compatibility.
In a numerical simulation, this means that the position of the respective boundary nodes of the fluid mesh can change.
In this work, the mesh deformation is considered to be induced by an impacting particle through surface erosion, but the overall
framework can handle both deposition and erosion.
Since the displacement of the boundary nodes has to be known a priori, a mapping from the particle's characteristics and
trajectory to this displacement is necessary, i.e., the erosion rate has to be known.
The erosion rate can be predicted by erosion models which correlate the dissipation energy of a particle upon an impact to the
erosion rate $\Delta Q$, such as in~\cite{Uzi2019}.
To convert this into a volume loss $\Delta V_p$, the physical density of the wall material $\rho_{\text{wall}}$ is used, leading to
\begin{align}
  \Delta V_p = \frac{\Delta Q \partmass}{\rho_{\text{wall}}}.
  \label{eq:theory:erosion_disp}
\end{align}
In Euler–Lagrange simulations, particles are treated as point-masses, which would physically result in a singular Dirac impulse; an unrealistic representation of wall deformation. To resolve this, particle impacts are modeled to induce a distributed displacement around the impact point, utilizing compactly supported RBFs $\psi: [0,\infty) \to \R$ to approximate realistic local deformation.
Following \citet{DeBoer2007}, the compactly supported RBF used here is defined as
\begin{align}
  \psi(\zeta)=\begin{cases}
    (1-\zeta)^4(4\zeta+1) &: 0 \leq \zeta \leq 1, \\
  0                     &: \zeta > 1,  \end{cases}
  \label{eq:theory:rbf_func}
\end{align}
where $\zeta=\zeta(\pos_{b_i},\pos_{b_j})=\norm{\pos_{b_i}-\pos_{b_j}}/R$ with $i,j=1\ldots,n_b$, $R \in \R^+$ being the support radius, $\pos_b$ the position of the boundary nodes, and $n_b$ the number of boundary nodes.
To prevent excessive mesh distortion, displacement is restricted to the wall-normal direction.
As such, the displacement $\mathbf{D}_{b_i}$ at a boundary node, $\pos_{b_i}$, in wall-normal direction is given as
\begin{align}
  \mathbf{D}_{b_i} = \sum_{j=1}^{n_p} \psi_p\lr{\norm{\pos_{b_i}-\mathbf{x}_{p_j}}} \gamma_i \normalvec,
  \label{eq:theory:dis}
\end{align}
where $\psi_p=\psi$ and $n_p$ is the number of impacting particles.
The weight vector $\gamma_i$ is calculated by matching the RBF to the node displacements of the impacting particle to ensure that the integral displacement of the boundary nodes is equal to the particle-induced deformation of a single particle which impacts at $\mathbf{x}_{p_j}$, i.e.,
\begin{align}
  \sum_{i=1}^{n_b} \psi_p\lr{\norm{\mathbf{x}_{p_j}-\pos_{b_i}}} \gamma_i V_{b_i} = \Delta V_{p_j},
  \label{eq:theory:dvolpart}
\end{align}
with $V_{b_i} = \w_{\text{vol}} \J_{b_i}$ being the volume of the boundary node, $\w_{\text{vol}}$ are the volume quadrature
weights, $\J_{b_i}$ is the determinant of the Jacobian, and $\Delta V_{p_j}$ is the volume loss imposed by the impacting particle coming from the particle erosion model, computed using~\cref{eq:theory:erosion_disp}.
As illustrated in~\cref{fig:theory:rbf_ale}, the RBF ensures that deformations are localized to the vicinity of the impact point, $\norm{\pos_{b_i}-\partpos} \leq R_{S_p}$ for $i=1,\dots,n_b$, and vanish elsewhere.
The support radius $R_{S_p} \in \R^+$ is defined as the maximum of the particle diameter and the local grid spacing; the latter
constraint ensures that at least one boundary degree of freedom is influenced by the impact.

Finally, since the continuous form of the mesh velocity $\meshVel$ is unknown, it is derived from the discrete node positions via numerical differentiation, leading to
\begin{align}
  \meshVel_{b_i} = \frac{\mathbf{D}_{b_i}}{\Delta t_{\meshVel}},
  \label{eq:methods:ale_vel}
\end{align}
with the time interval $\Delta t_{\meshVel}$ over which the deformation takes places.

To summarize, the particle-induced mesh deformation algorithm consists of the following two steps.
First, the calculation of the particle-induced volume loss using~\cref{eq:theory:dvolpart} and the mesh velocity used by the ALE
approach, cf.~\cref{eq:methods:ale_vel}.
Second, the displacement of the affected boundary nodes according to the volume loss using compact RBFs to induce a distributed
displacement around the impact point, cf.~\cref{eq:theory:dis}.
Further details are provided in~\cite{Schwarz2024}.

%% file: 3_numerical_methods.tex
\section{Numerical Methods and Implementation}%
\label{sec:methods}
In the following, we briefly discuss the numerical treatment of the governing equations for the fluid and dispersed phase, focusing
on the ALE method for surface deformation and the non-conforming sliding mesh interface.
The methods presented in this study are incorporated within the open-source framework~\elexi.

\subsection{Continuous Phase}%
\subsubsection{Discontinuous Galerkin Spectral Element Method}%
\label{sec:methods:dgsem}

The computational domain $\smash{\Omega_t \subseteq \mathbb{R}^3}$ is discretized by non-overlapping, (non-)conforming hexahedral elements with
six possibly curved element faces, approximated in a tensor product manner by one-dimensional Lagrange polynomials $\ell$ up to degree
$\ppngeo$.
The element-local solution $\cons = \cons(\refpos,t)$ is approximated by a tensor product of one-dimensional nodal
Lagrange basis functions of degree $\ppn$.
The entropy stable DGSEM on Legendre--Gauss--Lobatto nodes can be retrieved using the diagonal-norm summation-by-parts property, ${\Qm}^T+\Qm=\Bm$, and
an adequate two-point flux, leading to the following semi-discrete formulation
\begin{align}
  \Mm(\J \cons)_t = -& \sum\limits_{p=1}^3 (2 \Qm \circ \fcont^{(p)}) \mathbf{1} \nonumber \\
  +& \sum_{\zeta=1}^{6} \Vm_f^\transpose  \Wm_f [\J^{(\zeta)} (\fcont^{(\zeta)} \cdot \normalvec^{(\zeta)})^\ast - \fcont^{(\zeta)} \cdot
  \normalvecref^{(\zeta)}]
\end{align}
with the vector of all ones $\mathbf{1} \in \R^2$, $\Qm =  \Mm \Dm, \ \Bm = \mathrm{diag}(-1,0,\ldots,0,1)$, the polynomial derivative matrix $\Dm_{ij}=\ell_j'(\refpos_i)$, the mass matrix $\Mm =
\mathrm{diag}(\w_0,\ldots,\w_{\mathcal{N}})$ comprised of the quadrature weights $\w$, $\Vm_f$ being the Lagrange polynomials
evaluated at the corresponding surface, and $\Wm_f$ are the quadrature weights on the surface.
The outward pointing normal vector in physical space is $\normalvec$ and in reference space $\normalvecref$, and $\J^{(\zeta)} =
\abs{\lr{\text{adj}~\Jm}^{\transpose} \normalvecref^{(\zeta)}}$ is the surface element.
The contravariant fluxes $\boldsymbol{\mathcal{F}}$ are computed using an entropy-conservative flux
function~\cite{Chandrashekar2013} and an adequate state function $\qsharp$~\cite{Schnucke2018}, and, if not stated otherwise, the
numerical fluxes $(\fcont^{(\zeta)} \cdot \normalvec^{(\zeta)})^\ast$ are composed of the entropy-conservative flux and a Roe-type dissipation \cite{Roe1981} with the entropy fix by \cite{Harten1983}.
The viscous fluxes are computed using the BR1 scheme~\cite{Bassi1997}.

To mitigate oscillations of the high-order DGSEM at discontinuities, an element-wise convex blending procedure based on the work
of~\cite{Hennemann2021,Schwarz2025b} is utilized as a shock capturing method.
For this, a second-order total-variation diminishing FV subcell scheme is blended with the high-order DGSEM using the
blending coefficient $\FValpha \in [0,1]$, triggered by a modal troubled-cell indicator.
Note that surface fluxes and states must also be blended to account for general inconsistencies between the DG and FV grid velocities ($\meshVel_f^{FV} \neq \meshVel_f$) and solution states ($\cons_f^{FV}\neq\cons_f$).

An explicit low-storage fourth-order accurate Runge--Kutta scheme with $n_{\text{RK}}=5$ stages~\cite{Carpenter1994} is
employed for the integration in time, in accordance with the method of lines approach, with a relative Courant--Friedrichs--Lewy (CFL) number of $\text{CFL}=0.9$.
Its update is given as
\begin{align*}
  \cons_i  &= \cons_{i-1} + B_i d\cons_i, \ i = 0, ..., n_{\text{RK}}, \\
  d\cons_i &= A_i d\cons_{i-1} + \Delta t \cons_{t,i},
\end{align*}
with $\Delta t$ being the timestep, $A_i$ and $B_i$ are the low-storage equivalents of the Butcher variables.
For further details on DGSEM and applications, readers are referred to~\cite{Beck2014, Krais2019, Schwarz2025, Kurz2025, Keim2026}.

\subsubsection{Discrete Geometric Conservation Law}
Following~\cite{Minoli2011}, the GCL,~\cref{eq:theory:gcl}, is discretized by the same scheme as the considered conservation
equations to guarantee free-stream preservation on a discrete level,
The equivalent semi-discrete DGSE discretization for the GCL is given as
\begin{align}
  \Mm \J_t = & \sum\limits_{p=1}^3 (2 \Qm \circ \boldsymbol{\mathcal{V}}^{(p)}) \mathbf{1} \nonumber \\
  + & \sum_{\zeta=1}^{6} \Vm_f^\transpose  \Wm_f \lre{\J^{(\zeta)} (\boldsymbol{\mathcal{V}}^{(\zeta)} \cdot \normalvec) - \boldsymbol{\mathcal{V}}^{(\zeta)} \cdot \normalvecref},
  \label{eq:gcl_discrete}
\end{align}
where $\boldsymbol{\mathcal{V}} = \avg{\mathrm{adj}\Jm} \cdot \avg{\meshVel}$ is the arithmetic mean of the contravariant mesh
velocities, and $\boldsymbol{\mathcal{V}}^{(\zeta)}$ denotes the mesh velocities at the surface.
In order to guarantee non-overlapping elements for $t>0$, it is necessary that the mesh velocities are continuous across the cells and thus unique.

\subsubsection{Sliding Mesh Approach}
\begin{figure}[htbp]
  \centering
  \includegraphics[width=0.5\columnwidth]{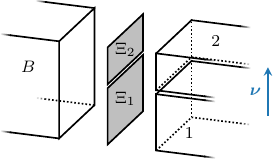}
  \caption{Non-conforming element interface between a big element ($B$) face and two small element ($1,2$) faces. The
  corresponding virtual mortar sides are denoted by $\Xi_{1,2}$.}
  \label{fig:methods:mortar}
\end{figure}

When introducing subdomains with different movement, the relative motion of the domain boundaries introduces nonconforming grids on either side of the interface.
The sliding mesh approach ensures conservative coupling of the flux exchange between the subdomains via the mortar method.
The mortar approach was originally proposed by~\citet{Mavriplis1989} and later extended to compressible
flows~\cite{Kopriva1996a,Kopriva1998}.
It preserves the mesh geometry within each subdomain by introducing virtual mortar sides $\Xi_{1,2}$ within the non-conforming interface such that each mortar side is exclusively connected to one element face on either side of the interface, cf.~\cref{fig:methods:mortar}.
To guarantee a conservative coupling, the integration within the discrete projection from the virtual mortar sides to the big side is decoupled from the general solution representation and always performed on Legendre--Gauss quadrature nodes.
Details on the implementation for the continuous phase are given in~\cite{Duerrwaechter2021}.

\subsubsection{Arbitrary Mesh Deformation}
For the time-accurate mesh deformation, the interpolation of the mesh displacement into the volume is required, which can be
accomplished using inverse distance weighting or, as in this paper, radial basis function interpolation.
While the evaluation of RBFs is more expensive than an inverse distance weighting, RBFs provide a better balance between efficiency and mesh quality.
The values of the RBF interpolation in each direction at $\pos_i$ are given as
\begin{align}
  \mathbf{s}(\pos_{i}) = \sum_{j=1}^n \hat{\mathbf{s}}_j \psi(\norm{\pos_{i}-\pos_{j}}) \hspace{0.2cm} \mathrm{s.t.} \hspace{0.2cm} s(\pos_{i})=\mathbf{D}(\pos_{i})
  \label{eq:methods:ale:rbf}
\end{align}
with the unknown coefficients $\mathbf{S} = \{\hat{\mathbf{s}}_j\}_{j=1}^n \in \R^{n \times 3}$, composed of the values of the interpolation
$s(\pos_{i})$ on each node $\{\pos_i\}_{i=1}^n$, i.e., the displacement $\mathbf{D}(\pos_{i})$ of the mesh nodes, with $n$ being the number of mesh nodes. Here, a deformation of the boundary points in normal direction, defined as $\mathbf{D}_b \in \R^{n \times 3}$, is considered, leading to
\begin{align}
  \mathbf{D}_b = \Vm_{\text{RBF}} \mathbf{S} \iff \mathbf{S} = \Vm_{\text{RBF}}^{-1} \mathbf{D}_b,
  \label{eq:methods:ale:rbf_matrix}
\end{align}
where $\mathbf{D}_{b_i} = \mathbf{D}(\pos_{b_i})$ denotes the displacement in each direction at $\{\pos_{b_i}\}_{i=1}^{n_b}$.
The Vandermonde matrix $\Vm_{\text{RBF}} \in \R^{n \times n}$ is comprised of the evaluation of the RBFs on each node,
$\Vm_{\text{RBF}}^{\{ij\}} = \psi(\norm{\pos_{i}-\pos_{j}}), \ i,j=1,\ldots,n$.
In this work, this displacement is computed using~\cref{eq:theory:dis} and the compactly supported RBF defined
in~\cref{eq:theory:rbf_func} is employed.

Determining $\mathbf{S}$ requires inverting $\Vm_{\text{RBF}}$ (cf. \cref{eq:methods:ale:rbf_matrix}), an $\mathcal{O}(n_b^2)$
operation that becomes prohibitive if the RBF is constantly rebuilt globally to accommodate load balancing or boundary node drift.
While recent methods rely on state-of-the-art surface and volume point reduction schemes to manage grid deformation overhead~\cite{Morelli2021}, our approach utilizes localized RBFs. Using compactly supported RBFs ensures the interpolation matrix remains sparse, allowing for local grid construction exclusively using mesh points within the support radius. To minimize computational overhead, an MPI-3 shared-memory parallelization strategy \cite{Kopper2022} is employed, significantly reducing communication and memory consumption.

\subsection{Dispersed Phase}%
\begin{figure}[!tb]
  \centering
  \includegraphics[width=0.75\columnwidth]{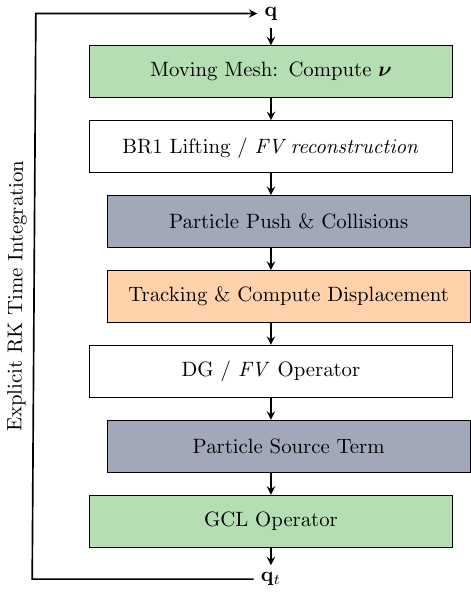}
  \caption{Flow chart of the DG operator for 4-way coupled particle-laden flow via the Euler--Lagrange approach. DG operations are shaded in white.
    Particle operations are shaded in gray with the particle routines affected by the moving mesh method highlighted in orange and the
  moving mesh operator of the continuous phase in green. See \cite{Kopper2023,Schwarz2025a} for a more detailed breakdown.}
  \label{fig:implementation:latency}
\end{figure}

\label{sec:methods:tracking}
The particle operator in each stage of the time stepping algorithm is depicted in \cref{fig:implementation:latency}, including the
particle interpolation, time integration, and tracking on arbitrarily moving meshes and across non-conforming sliding mesh interfaces.
More details on the numerical handling and implementation of the dispersed phase are given in~\cite{Kopper2023,Schwarz2025a}.

\subsubsection{Particle Interpolation, Time Integration and Tracking}%
At each stage, the particle push is determined which includes the following steps.

First, the localization of the particle in reference space, which is facilitated by a Cartesian background mesh that maps physical
coordinates to overlapping computational elements \cite{Kopper2022}.
For each candidate element, the particle’s reference coordinates $\partrefpos$ are resolved by finding the root of $\partpos - \boldsymbol{\chi}(\partrefpos) = 0$ via Newton's method.
The valid host element is uniquely identified when the solution satisfies $\partrefpos \in [-1, 1]^3$.

Second, the conserved variables are interpolated to the particle location. For DG or blended DG cells ($\FValpha <
\FValpha_\text{max}$), this is performed using the high-order polynomial expansion, given as
\begin{align}
  \cons(\partrefpos,t) = \sum_{i,j,k=0}^\ppn \hat{\cons}_{ijk}(t) l_i(\xi_p^1) l_j(\xi_p^2) l_k(\xi_p^3).
\end{align}
Conversely, for cells utilizing the second-order FV subcell scheme, the operation reduces to a linear interpolation of the conserved variables.

Third, the calculation of the corresponding force on the discrete particle, and the subsequent integration of the particle
trajectories in time using the updated particle acceleration.
The particle path within a single RK stage, $t \in [t^{n},t^{n+1}]$, is approximated using linear segments, defined as
\begin{align}
  \partpos(t; \alpha) = \partpos(t^{n}) + \alpha \frac{\parttrajectory}{\vert \parttrajectory \vert},\ \parttrajectory =
  \partpos(t^{n+1}) - \partpos(t^{n}),
  \label{eq:methods:path}
\end{align}
where $\parttrajectory$ represents the linear displacement vector and $\alpha \in [0, \vert \parttrajectory \vert], \ $ denotes the scalar distance along this segment.
To ensure temporal synchronization between the dispersed and continuous phases, the particles are advanced using the same RK scheme as the fluid solver.

Fourth, the final particle path is determined by tracking intersections with element boundaries and applying the relevant boundary conditions.
For curvilinear boundaries, intersections are identified using a dimension-reduction approach based on Bézier clipping and de Casteljau subdivision.
Each element side is represented as a Bézier polynomial surface, yielding
\begin{align}
  \bezierpoly(\xi,\eta) = \sum_{m=0}^{\ppngeo} \sum_{n=0}^{\ppngeo} \bezierdofs_{mn}
  \bezierbasis_m(\xi) \bezierbasis_n(\eta),
\end{align}
where $\bezierdofs$ are the Bézier control points and $(\xi,\eta) \in [-1,1]^2$ the side coordinates in reference space.
An intersection is confirmed by finding a root such that $\smash{\partpos(t;\alpha) \overset{!}{=} \bezierpoly(\xi,\eta)}$.
Computationally efficient linear or bi-linear solvers are substituted for planar faces, see~\cite{Ortwein2019} for further details. To maintain geometric watertightness, non-conforming interface intersections are evaluated exclusively on the smaller element faces \cite{Kopper2023}.

\subsubsection{Sliding Mesh Approach}
\label{sec:methods:disp_SM}
\begin{figure}[htbp]
  \includegraphics[width=\columnwidth]{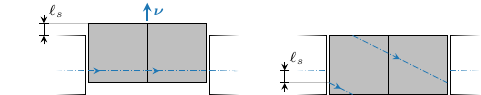}
  \caption{Particle path in the stationary (left) and relative (right) frame of reference of the continuous phase with $\ell_s$
  being the displacement and $\meshVel$ the mesh velocity.}
  \label{fig:methods:sm}
\end{figure}

As the tracking of the discrete phase is conducted in physical space via Bézier clipping, a continuous update of the Bézier
polynomials to account for mesh motion would introduce significant computational overhead at each time step.
To mitigate this overhead, tracking of the discrete phase within the moving mesh region is instead performed in the relative frame
of reference of the respective moving subdomain, where the crossing of a sliding mesh interface then corresponds to applying an instantaneous displacement of the particle path at the interface location and subsequent application of virtual forces.
The algorithmic treatment of an intersection of a particle with the sliding mesh interface parallels then that of internal mortar
sides, differing by the change in the frame of reference for the dispersed phase required when a particle traverses an interface.
This approach is illustrated in~\cref{fig:methods:sm}, which contrasts the particle path as observed from the stationary
frame of reference of the continuous phase with the corresponding trajectory from the relative frame of reference.

Upon identifying an intersection with the big mortar side, cf.~\cref{fig:methods:mortar}, the instantaneous displacement of the
small mortar sides at time $t^n$ is evaluated.
Since the particle path given in~\cref{eq:methods:path} assumes that the mesh position remains unchanged in
$t\in[t^n,t^{n+1}]$, the particle is virtually moved to the intersection point on the sliding mesh interface
using~\cref{eq:methods:path} and adjusted by the displacement $\ell_s(t) = \meshVel t - \ell_s(t=t_0)$, representing the physical
translation of the moving sub-domain.
It is evident that, owing to the tracking in the relative frame, the particle velocities require correction by the mesh velocity. This results in a discrete jump in the particle velocity as it passes across a sliding mesh interface.
Consequently, the RK registers corresponding to the intermediate stages must be modified to account for the frame transformation.
The corrected particle velocity in the $n$-th RK stage, $d(\partvel)_n = d(\partvel)_n + \gamma_n$ incorporates a correction factor
$\gamma_n$ derived from the following recursive relation
\begin{align}
  \gamma_j = -\meshVel - A_j \gamma_{j-1}, \quad \text{for } j = 2, \dots, n_{\text{RK}}, \ \gamma_1 = -\meshVel.
\end{align}
Further details are provided in~\cite{Kopper2024}.

\subsubsection{Arbitrary Mesh Deformation}
\label{sec:methods:disp_ALE}
For arbitrary mesh deformations away from the sliding mesh interface, particles are tracked in an absolute frame of reference. This inherently accounts for the mesh movement, providing correct particle trajectories without the need for further modifications. Although rebuilding the Bézier polynomials after each mesh deformation introduces a higher computational cost, it is strictly necessary to guarantee the geometric fidelity required for accurate particle tracking on meshes with arbitrary deformations.

%% file: 5_validation.tex
\section{Validation}%
\label{sec:validation}
Prior to presenting some application cases, the individual building blocks for particle-laden flow on moving meshes are validated.
These include the particle push in a compressible fluid, the spatial and temporal convergence for the two moving mesh methods, and
the local RBF interpolation.
For further details on the validation of the particulate phase, such as convergence properties for the static domain, and the
correctness of the one-, two-, and four-way coupling, the reader is referred to~\cite{Kopper2023,Schwarz2025a}.
A comprehensive validation of the continuous phase, the ALE method, and the sliding mesh approach is provided in~\cite{Krais2019,Duerrwaechter2021,Schwarz2025b}.
In this section, it is assumed that particle collisions are purely elastic, i.e., $\cor{n}=1$.
Under these assumption, the tangential component of the impulse is zero and the normal component reduces to $\smash{J_n = - 2 \partmass[r] (\partvel[r]^{(0)} \cdot \normalvec_p)}$.
The high-order DGSE/FV subcell scheme is used for the spatial discretization of the continuous phase, which is assumed to be inviscid, i.e., only the Euler equations are considered in this section, if not stated otherwise.
All meshes in this paper are generated using the open-source pre-processing framework pyHOPE~\cite{Kopper2025}.
It is imperative to acknowledge that all quantities are non-dimensional.

\subsection{Particle Push}
\begin{figure}[htbp]
  \centering
  \includegraphics[width=\columnwidth]{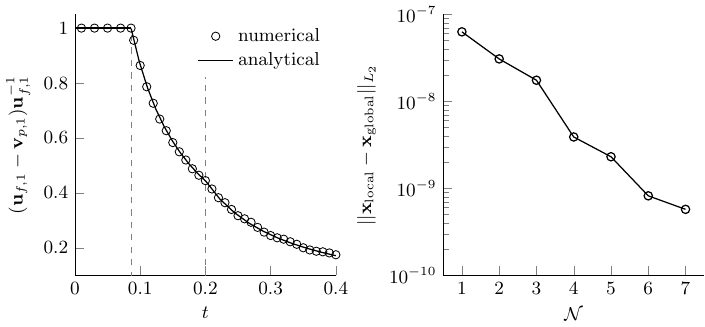}
  \caption{Left: Normalized relative particle velocity for the Sod shock tube. Right: Validation of the local RBF interpolation. $L_2$-error of the mesh node positions for {\usefont{U}{BOONDOX-cal}{m}{n}p}-refinement compared to its global counterpart.}
  \label{fig:validation:defectv_rbf}
\end{figure}

Following~\cite{Kaiser2021}, the validity of the particle push in a compressible context is assessed via a single particle with
$\fluidvel=C$, $C = \smash{\frac{3 C_D \fluiddens}{4 \partdens \partdiam}}$, and the drag coefficient, $C_D = 1$, moving in a Sod shock tube.
The assumption that $\fluidvel=C$ is only valid if the underlying fluid solution is piecewise constant, i.e., not for the rarefaction wave.
Thus, the normalized relative velocity can be defined as
\begin{align}
  \frac{\urel}{\fluidvel} &= \lr{C \lr{t-t_0} \fluidvel + \frac{\fluidvel}{\urel|_{t=t_0}}}^{-1}.
\end{align}
The relative time $t-t_0$ and the relative particle velocity $\urel|_{t=t_0}$ are chosen in accordance to the considered characteristic wave, i.e.,
$t-t_0=2-0.086$, $\urel|_{t=t_0}=\fluidvel$ and $t-t_0=0.4-0.2$, $\urel|_{t=t_0}=\partvel(t=t_0)-\fluidvel$ for the shock wave and the contact discontinuity, respectively.
The particle is initialized at $\partpos|_{t=t_0}=\arr{0.65,0,0}$, to the right of the initial discontinuity, with the local fluid velocity and $\partdens=10$, $\partdiam=0.001$.
The intersection with the shock wave at $t=0.086$ results in an acceleration of the particle, which is again visible for the contact
discontinuity at $t\approx 0.2$, as depicted in~\cref{fig:validation:defectv_rbf} (left).
The kink in the normalized relative velocity can be attributed to the jump in density, $\fluiddens$, over the contact discontinuity.
Altogether, the analytical and numerical results are in excellent agreement.

\subsection{Convergence Tests}
The following test cases validate the spatial and temporal convergence of the moving mesh approaches for the continuous and the discrete phase
under mesh refinement ({\usefont{U}{BOONDOX-cal}{m}{n}h}-convergence), refinement of the polynomial solution
({\usefont{U}{BOONDOX-cal}{m}{n}p}-convergence), and refinement of the time step ({\usefont{U}{BOONDOX-cal}{m}{n}t}-convergence).

\subsubsection{Arbitrary Lagrangian--Eulerian Approach}
\begin{figure}
  \includegraphics[width=\columnwidth]{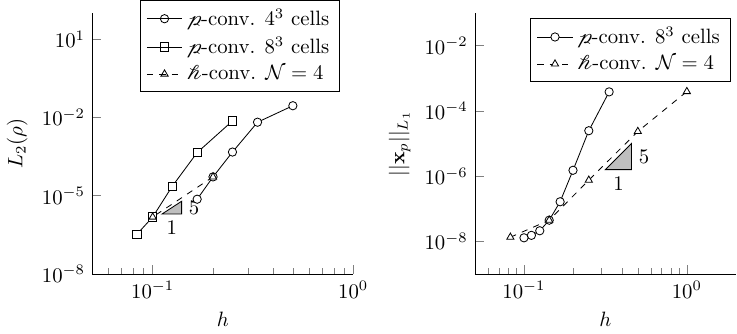}
  \caption{Experimental rate of convergence for {\usefont{U}{BOONDOX-cal}{m}{n}p}-refinement and
  {\usefont{U}{BOONDOX-cal}{m}{n}h}-refinement for the continuous (left) and discrete (right) phase with the arbitrary Lagrangian--Eulerian approach on a conformal grid.}%
  \label{fig:validation:ale}
\end{figure}

In order to validate the ALE approach for the continuous phase, the computational domain $\Omega_t \in [0, 4]^3$ is tessellated into $4^3$/$8^3$ hexahedral elements with fully periodic boundary conditions and the mesh is deformed sinusoidally in time~\cite{Minoli2011}.
Initial conditions are chosen according to~\cite{Gassner2009a}, prescribing an oblique sine wave with unit frequency and amplitude $a=0.1$.
Experimental rates of convergence (EOC) for {\usefont{U}{BOONDOX-cal}{m}{n}p}-refinement and {\usefont{U}{BOONDOX-cal}{m}{n}h}-refinement are shown in \cref{fig:validation:ale} (left) with the EOC confirming the expected spectral and geometric convergence of the error, respectively.

\begin{figure}[htbp]
  \includegraphics[width=.99\columnwidth]{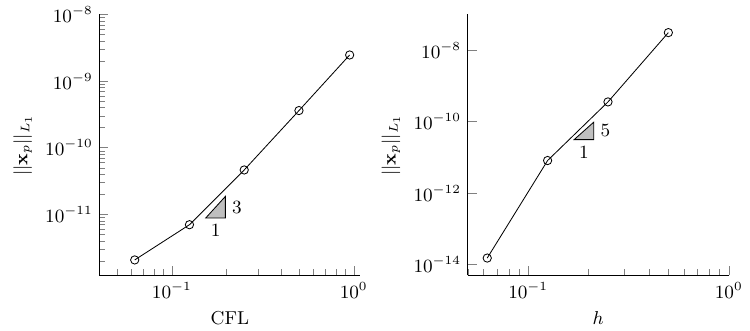}
  \caption{Experimental rate of convergence for {\usefont{U}{BOONDOX-cal}{m}{n}t}-convergence for the discrete phase with $CFL =
    \{0.0625\cdot2^k:k\in\N,k\in[0,4]\}$ and $\ppn=6$ (left) and {\usefont{U}{BOONDOX-cal}{m}{n}h}-convergence for the discrete
  phase interacting with a moving curved face with $\ppngeo=\ppn=4$ (right).}
  \label{fig:validation:ale:thgeoconv}
\end{figure}

Validation for the discrete phase is performed on a cubical domain $\Omega_t \in [-1, 1]^3$, with the test case conditions given in~\cite{Kopper2023}.
A single particle is emitted at $\partpos|_{t=0} = \arr{-0.95, 0.95, 0}$ with an initial velocity of $\partvel|_{t=0} = \arr{0,
0.01, 0}$ and integrated in time until $t=1$ while the mesh is subjected to the same sinusoidal motion as for the continuous phase.
The EOC for the $L_1$ error of the particle position at the end of the simulation are depicted in \cref{fig:validation:ale} (right) for the spatial convergence study and in~\cref{fig:validation:ale:thgeoconv} (left) for the temporal convergence study.
In all cases, the EOC confirms to the expected slope until numerical errors in the {\usefont{U}{BOONDOX-cal}{m}{n}h}-convergence
study limit the accuracy due to the time step size.

\begin{figure}[htbp]
  \centering
  \includegraphics[width=.3\columnwidth]{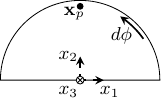}
  \caption{Setup of the convergence test for curved faces. \label{fig:validation:half_circle}}
\end{figure}

Since only straight faces have been investigated so far, the accuracy of intersections of particles with moving curved faces is now evaluated.
For this, a half circle with a radius of $r_C=2$ and a perimeter of $l_C=4\pi$ in the $xy$-plane is rotated around the $z$-axis
with $d\phi=0.1^\circ$ at each time step, cf.~\cref{fig:validation:half_circle}.
A single particle is placed in a quiescent fluid with $\rho=p=1$ at $\partpos|_{t=0}=\arr{0,1.4,0}$ and moved until $t=0.6$ with $\partvel|_{t=0}=\arr{1.2,1.2,0}$.
The numerical intersection point of the particle with the circle is compared to the analytical one, and the discrete $\text{L}_1$ error
for different grid resolutions with a solution and geometric polynomial order of $\ppn=\ppngeo=4$ is chosen as error metric.
The results in~\cref{fig:validation:ale:thgeoconv} (right) demonstrate that the optimal convergence rate of $\ppngeo+1$ is reached for all grid
resolutions.

\subsubsection{Sliding Mesh Interface}
\begin{figure}
  \includegraphics[width=\columnwidth]{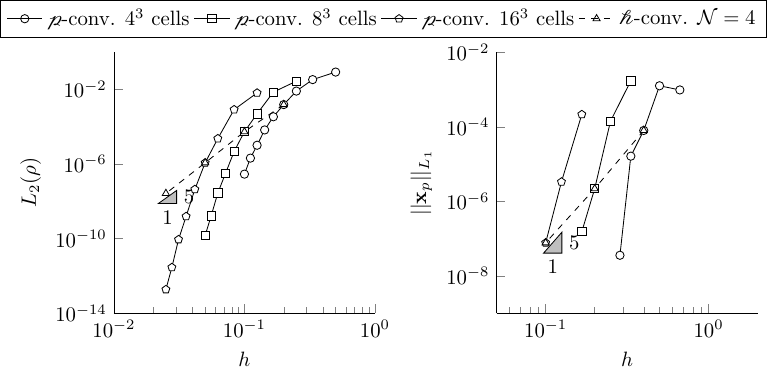}
  \caption{Experimental rate of convergence for {\usefont{U}{BOONDOX-cal}{m}{n}p}-refinement and {\usefont{U}{BOONDOX-cal}{m}{n}h}-refinement for continuous (left) and discrete (right) phase with the arbitrary Lagrangian--Eulerian approach on a nonconforming grid with two sliding mesh interfaces.}%
  \label{fig:validation:sm}
\end{figure}

Validation of the sliding mesh approach is performed using the identical setup as for the conforming grid test case.
However, for the validation of the continuous phase, the central subdomain $x \in [1,3]$ is translated with a constant grid velocity
vector $\meshVel = \arr{0, 1, 0}$ while the outer subdomains remain invariant.
Again, EOC for {\usefont{U}{BOONDOX-cal}{m}{n}p}-refinement and {\usefont{U}{BOONDOX-cal}{m}{n}h}-refinement are shown in \cref{fig:validation:sm} (left) with the EOC confirming the expected spectral and geometric convergence of the error, respectively.

Similarly, the validation for the discrete phase is performed on the cubical domain $\Omega_t \in [-1, 1]^3$.
The moving region is chosen as the central subdomain $x \in [-0.5, 0.5]$ and translated with a constant grid velocity vector
$\meshVel = \arr{0, 1, 0}$ while the other subdomains remain at rest.
The EOC for the $L_1$ error of the particle position at the end of the simulation are depicted in \cref{fig:validation:sm} (right).
Once more, the EOC confirm to the expected slope once the sinusoidal solution of the continuous phase is sufficiently approximated by the polynomial representation.
The EOC for both {\usefont{U}{BOONDOX-cal}{m}{n}t}-convergence and {\usefont{U}{BOONDOX-cal}{m}{n}h}-convergence of the discrete phase along a moving curved interface was established in the preceding section.

\subsubsection{RBF Interpolation}
To validate the accuracy of the local RBF implementation, its performance is evaluated against a global RBF baseline. A particle-induced mesh deformation is simulated on a domain $\Omega_t \in [-1,1]^3$ featuring $8^3$ elements with $\ppn=4$.
A single particle ($\partpos|_{t=0} = \arr{0.55, 0.55, 0.05}$) is evolved until $t=0.6$ on a uniform flow field using $R_S = R_{S_p} = 0.3$.
The resulting mesh deformations for different polynomial degrees $\ppn$ are compared in \cref{fig:validation:defectv_rbf} (right). The $L_2$
error between the locally and globally build RBFs is lower then $\mathcal{O}(10^{-7})$, confirming implementation consistency.
Furthermore, an $L_2$ error analysis across varying $\ppn$ illustrates that the mesh velocity interpolation error decreases with increasing polynomial degree, with convergence rates determined by the chosen support radius.

%% file: 7_application.tex
\section{Applications}%
\label{sec:application}

This section demonstrates the applicability of the \elexi framework to complex, large-scale applications with moving domains.
The ALE approach is applied to highly challenging particle-laden flow around a supersonic compressor blade to predict the resulting surface erosion.
To demonstrate the sliding mesh approach, the framework is utilized to simulate the computationally demanding particle-laden flow within a similar transonic compressor cascade experiencing upstream wakes from a cylindrical generator.
Although a unified simulation simultaneously deploying both methodologies is fully supported by the underlying architecture, these two moving mesh algorithms are evaluated independently here, as a joint application falls outside the scope of the present study.
These cases serve as numerical demonstrations; consequently, detailed physical investigations of the flow phenomena are beyond the scope of this study.

\subsection{Particle-Wall Deformations}%
\label{application:ale}
An infinite rectilinear compressor cascade with the uneroded NASA Rotor 37 profile~\cite{Reid1978,Moore1978} at $50\%$ span is utilized for the current investigations.
The rotor operates at design speed with an inflow Mach number of $M=1.4$, $\rho_1=\SI{0.968}{\kilogram \meter^{-3}}$, $p_1=\SI{72722}{\pascal}$, and a Reynolds number of $Re=\num{1472525}$, based on the chord length $c=\SI{0.0557}{\meter}$.
The computational domain around a single blade $\Omega_t=[-0.23677,0.24702]\times[-0.03576,0.08478]\times[0,0.00278]\ [\si{m}]$ is tessellated into
\num{1202760} elements with $\ppn=\ppngeo=4$ and periodic boundary conditions in $y$- and $z$-directions.
Supersonic inflow conditions and a pressure outflow condition, $p_2 = \SI{136900}{\pascal}$, are prescribed in $x$-direction (left to right), while adiabatic, no-slip conditions are imposed on the surface of the rotor blade.
Wall-normal spacing is selected to obtain a non-dimensional wall distance of $y^+ \leq \num{1.59}$, thus conforming to a wall-resolved LES.
The hybrid DGSE/FV subcell scheme is utilized for shock capturing with $\FValpha \in [0.01,0.5]$, while for
$\FValpha<0.01$, a pure DG and for $\FValpha>0.5$ a pure FV subcell method is employed.
The characteristic flow timescale is defined as $\characflowtime=\characlength u_2^{-1} \approx 9.62740~\eh{-5}\si{s}$ with
the characteristic length $\characlength = 0.0377\si{m}$ given as the cascade pitch and $u_2=\SI{391.409}{\meter \per \second}$ as the inflow
velocity in radial direction.
The domain is extended at the beginning and at the end by a sponge zone with grid stretching to prevent an interaction of the bow
shock with the inlet boundary condition and to dampen turbulent structures near the outflow, respectively.
Spherical quartz particles with a density of $\rho_p = 2650\text{kg m}^{-3}$ are introduced at a distance of \num{1.9} axial chord
lengths upstream from the leading edge, with their initial velocities set to match the instantaneous fluid velocity. The particle diameters
are fixed to specific values spanning a Stokes number range of $\text{St} \in [0.001, 100]$, $\smash{\text{St} = \frac{\rho_p d_p^2
\mathbf{u}_f}{18 \mu_f c}}$, utilizing identical emission probabilities for each Stokes number.

\begin{figure}
  \includegraphics[width=\columnwidth]{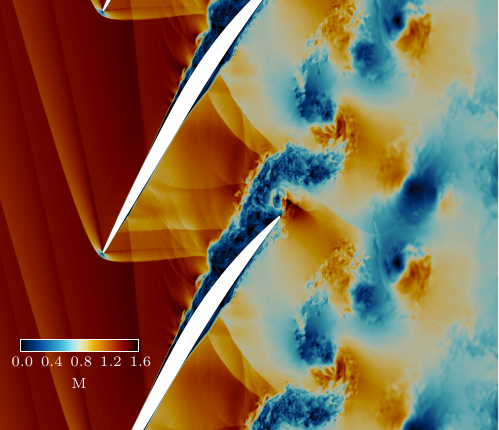}
  \caption{Instantaneous flow field colored by the Mach number around the compressor blade at design speed and $312\characflowtime$.}
  \label{fig:applications:ale:flowfield}
\end{figure}

The instantaneous flow field is visualized in \cref{fig:applications:ale:flowfield}. A detached bow shock is observed at the leading
edge of the rotor, followed by Prandtl--Meyer expansion waves and a complex passage shock wave system. Laminar boundary layers are
visible on both the pressure and suction sides, transitioning to the turbulent regime near the mid-chord. This transition interacts with the passage shock waves, resulting in a highly unsteady flow system.

To model the particle-induced deformation of the blade's surface, the boundary nodes of the blade are continuously deformed during
the simulation according to the algorithm presented in~\cref{sec:theory:partdeform}.
There is no mesh deformation in the third (spanwise) direction and the displacement is identical for all boundary nodes which
share the same $z$-position, since the cascade is assumed to represent the blade-to-blade plane.

\begin{figure}
  \includegraphics[width=\columnwidth]{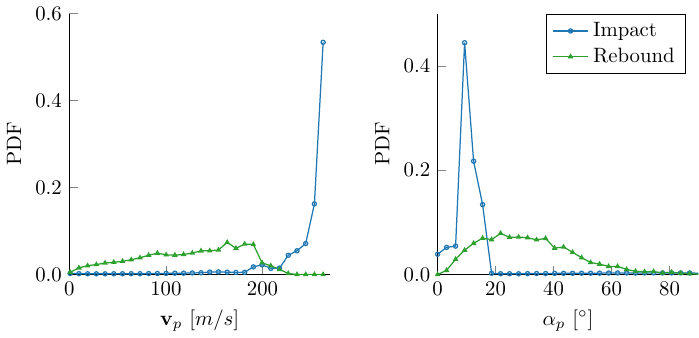}%
  \caption{Probability distribution function (PDF) of particle impact and rebound along the blade's surface predicted by the LES for the particle velocities (left) and impact/rebound angle (right).}
  \label{fig:applications:ale:rebound}
\end{figure}

The particle impact and rebound distribution along the blade's surface, as shown in \cref{fig:applications:ale:rebound}, highlights the distinct kinematics driven by the empirical rebound model~\cite{Schwarz2022}. Prior to impact, the initial particle emission conditions restrict the incoming particles to a highly concentrated range of velocities and impact angles, $\alpha_p$. Upon collision, however, the utilized rebound model introduces a pronounced scattering effect, significantly broadening the distribution of both particle velocities and reflection angles post-rebound.

\begin{figure}
  \includegraphics[width=\columnwidth]{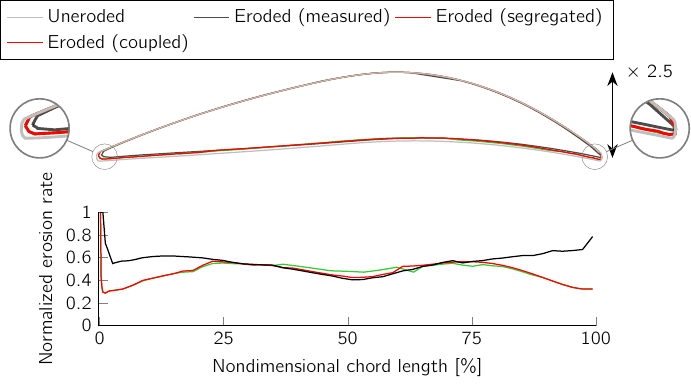}%
  \caption{Sketch of the rotor blade deformed by the ALE approach compared to the measured, eroded rotor blade given in~\cite{Lorenz2022,Hartmann2022}. The solid, grey line indicates the uneroded blade profile. For reasons of clarity, in the present sketch, the rotor height is scaled by a factor of around two compared to the chord length.}
  \label{fig:applications:ale:erosion}
\end{figure}

To quantify the structural impact of these dynamics, the local material degradation along the blade surface is evaluated.
The resulting surface erosion rate of the compressor blade normalized by the maximal erosion rate along the chord length is illustrated in \cref{fig:applications:ale:erosion}.
Experimental data for measured surface variations due to particle erosion on the NASA Rotor 37 is adapted from Lorenz et al.~\cite{Lorenz2022}.
Thus, this experimental data serves as reference for this validation case.
The numerical results demonstrate strong overall agreement with the measured rotor blade deformation, exhibiting only minor deviations near the leading and trailing edges in this complex multi-physis application.
The slight underprediction near the leading and trailing edges stems from a fundamental limitation in the underlying empirical framework. The utilized rebound and erosion models are traditionally derived from experimental data gathered using flat-plate specimens under uniform impact angles. Consequently, they lack the geometric fidelity required to inherently account for the highly localized curvature and strong local flow deceleration characteristic of the leading and trailing edges.
In these regions, particularly at the leading edge, the rapid deflection of the continuous phase creates a localized shielding effect, turning smaller particles away before impact, while the steep, continuously varying surface curvature alters the effective localized impact angles in a manner not fully captured by flat-plate empirical correlations.

\begin{figure}
  \includegraphics[width=\columnwidth]{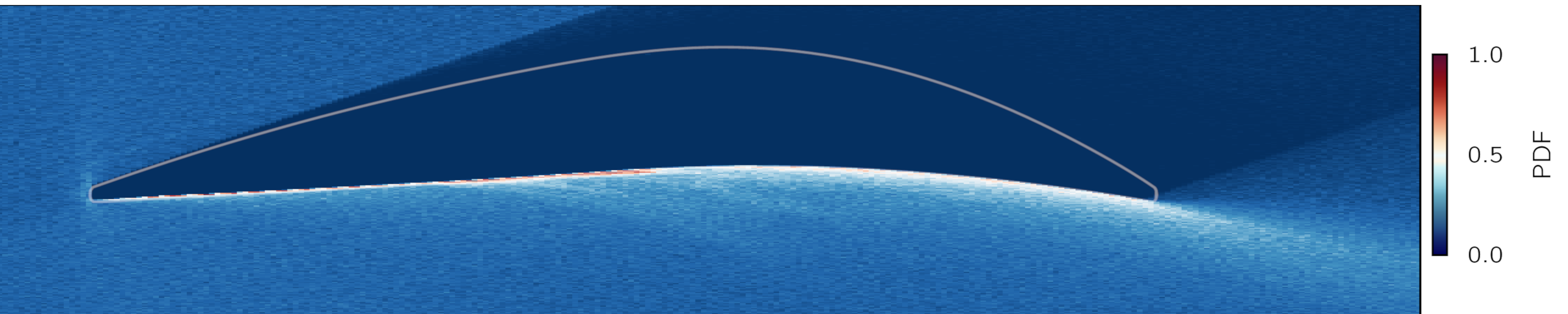}
  \caption{Normalized, two-dimensional particle density distribution in the $x_1 x_2$ - plane over the last 32$\tau^\ast$. The rotor
  blade is highlighted in gray. The uneroded rotor blade is illustrated and the particles are reflected at the blade’s surface with
$\Delta \mathbf{J}_p \neq \Null$.}
  \label{fig:applications:ale:partdens}
\end{figure}

The influence of these impacts on the particle dynamics is further highlighted in \cref{fig:applications:ale:partdens}, which depicts the normalized particle distribution in the $xy$-plane, averaged over the spanwise ($z$) direction for the final $32$ characteristic flow times. Crucially, this distribution reveals a massive particle concentration localized along the pressure side, perfectly mirroring the high-wear zone identified in the erosion profiles. Furthermore, the visualization clearly captures the non-symmetric scattering of the particle cloud post-impact, a direct physical consequence of the asymmetric reflection angles dictated by the empirical rebound model.
Ultimately, these altered trajectories propagate past the trailing edge, heavily reshaping the particle distribution profile in the blade's wake.

In summary, the coupled Euler--Lagrange approach successfully captures the non-linear interaction between the compressible flow features and the particulate phase. By accurately predicting both the high-velocity shock system and the resulting material degradation, the framework demonstrates its capability to resolve long-term surface evolution in complex, moving geometries.
Furthermore, numerical analysis provided insight into the highly complex particle trajectories and interactions, which are inaccessible during standard erosion cascade experiments.

\subsection{Cascade with Periodic Wakes}%
\label{application:sliding}
As second application case, the developed framework is applied to a transonic compressor cascade, subjected to temporally periodic wakes.
Using the same cascade geometry as in \cref{application:ale}, cylindrical wake generators are added upstream of the blade row at a distance of \num{0.65} axial chord lengths from the leading edge.
A sliding mesh interface is placed between the wake generators and the blade row to account for the relative motion.
The wake generators are meshed as fully-structured O-grid, adding \num{83640} additional cells for a total of \num{1286400} hexahedral mesh elements.
Wall-normal spacing is selected to obtain a non-dimensional wall distance of $y^+ \leq \num{1.4}$, thus conforming to a wall-resolved LES.
The selected operating point corresponds to a near-stall flow condition at \SI{60}{\percent} design speed at a relative inlet Mach
number of $M=\num{0.824}$, $\rho_1 = \SI{1.173}{\kilogram \meter^{-3}}$ and $p_1 = \SI{95 262}{\pascal}$, and a Reynolds number of
$Re=\num{972550}$, assuming sea-level idle conditions.
Subsonic inflow conditions and a pressure outflow condition, $p_2 = \SI{113 332}{\pascal}$, are prescribed in $x$-direction (left
to right), while adiabatic, no-slip conditions are imposed on the surface of the rotor blade and the cylindrical wake generator.
The discrete phase is initialized in a similar manner as in~\cref{application:ale}.

\begin{figure}
  \includegraphics[width=\columnwidth]{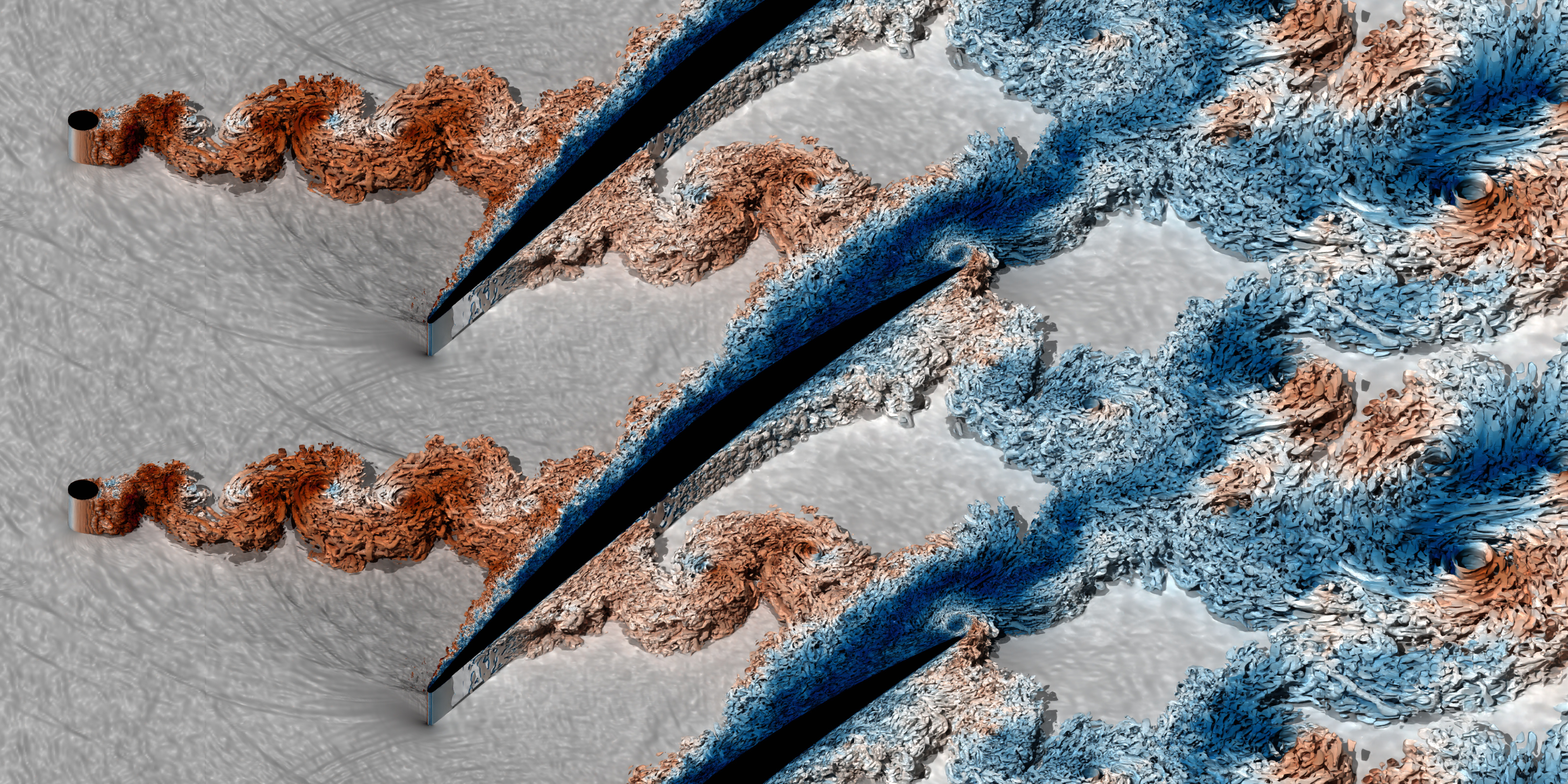}%
  \caption{Instantaneous isocontour of Q-criterion colored by Mach number around the compressor blade including the wake generator with Schlieren visualization in the background.}%
  \label{fig:applications:sliding:qcrit}
\end{figure}

A visualization of the instantaneous isocontour of the Q-criterion colored by Mach number is shown in \cref{fig:applications:sliding:qcrit}.
Clearly visible is the vortical wake of the cylindrical wake generators which enters the passage at mid-pitch for the depicted point in time.
The wake interacts with the transonic region near the leading edge and the detachment region downstream on the suction side while experiencing only minor wake bending while traversing the passage.
Once the fluid exits the passage, the wake-wake interactions results in temporally varying vortex amplification and attenuation.
The Schlieren visualization in the background indicates acoustic disturbances moving upstream and interacting with the generated wakes.

\begin{figure}
  \includegraphics[width=\columnwidth]{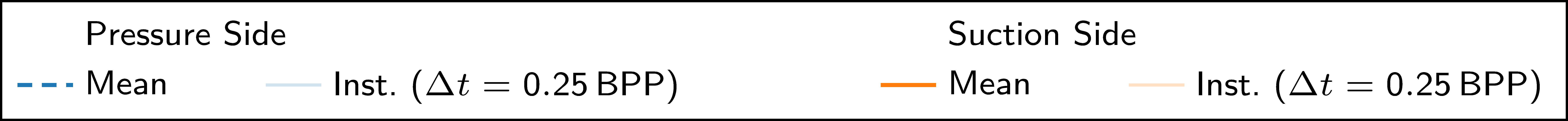}\\%
  \includegraphics[width=.5\columnwidth]{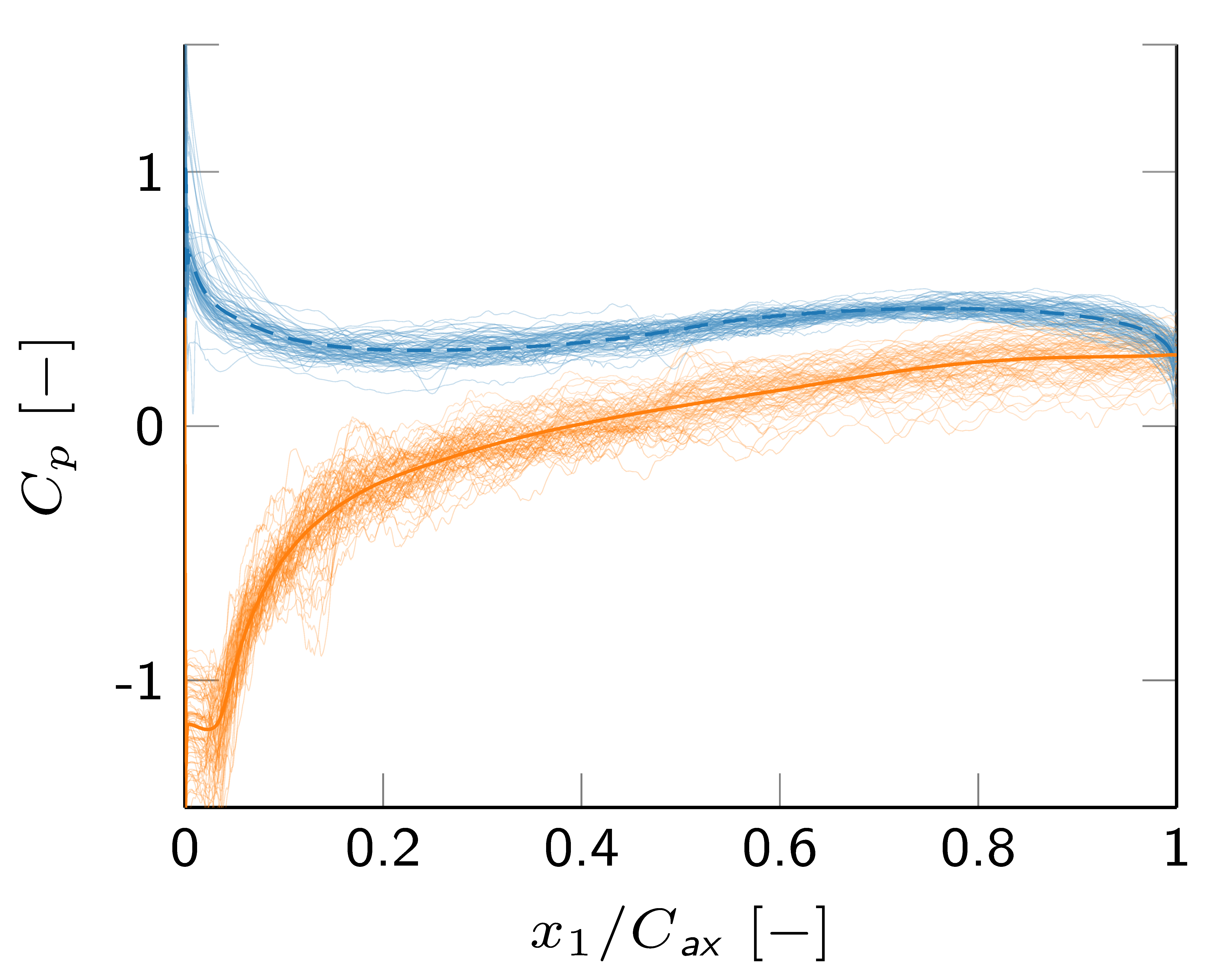}%
  \includegraphics[width=.5\columnwidth]{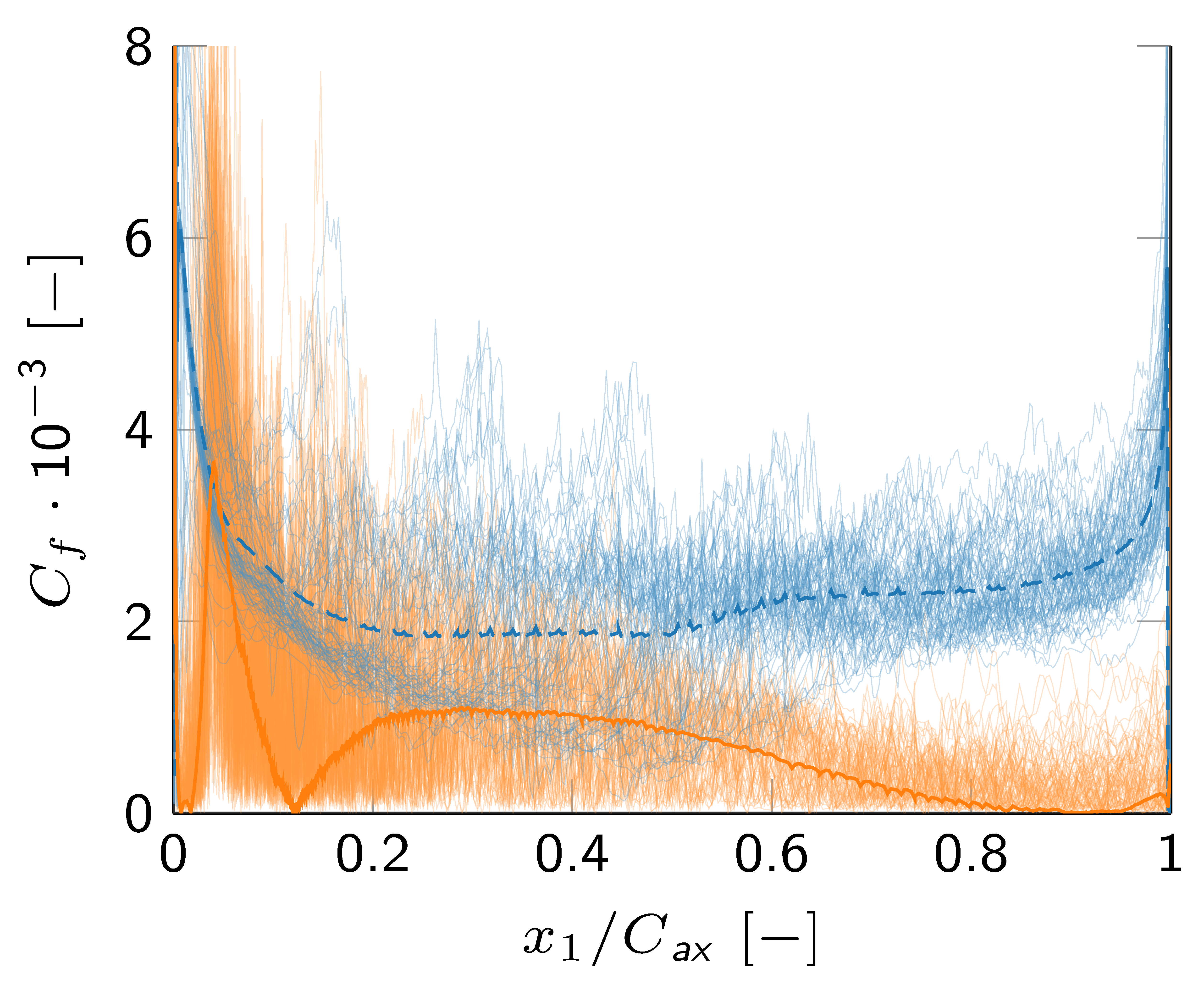}%
  \caption{Distribution of instantaneous and mean pressure coefficient $C_p$ and skin friction coefficient $C_f$ against relative position along the axial chord $x_{ax}/C_{ax}$.}%
  \label{fig:applications:sliding:wall}
\end{figure}

Instantaneous and time-averaged distributions of the pressure coefficient $C_p$ and the skin friction coefficient $C_f$ are depicted in \cref{fig:applications:sliding:wall}.
After the initial transient, statistics are gathered for 20 blade passing periods (BPPs).
As expected from the low operating point, overall pressure difference between suction and pressure side and consequently the work performed by the rotor stage remains small.
The skin friction coefficient distributions reveals a bi-modal pattern on the pressure side as a thin separation region is periodically re-energized by the wake passage.
This boundary layer re-energization facilitates the formation of a transient calmed region as the flow field gradually relaxes back towards a laminar boundary layer on the verge of separation.
Naturally, these localized variations in boundary layer dynamics and fluid velocity directly modulate the transport and acceleration of the dispersed phase.

\begin{figure}
  \includegraphics[width=\columnwidth]{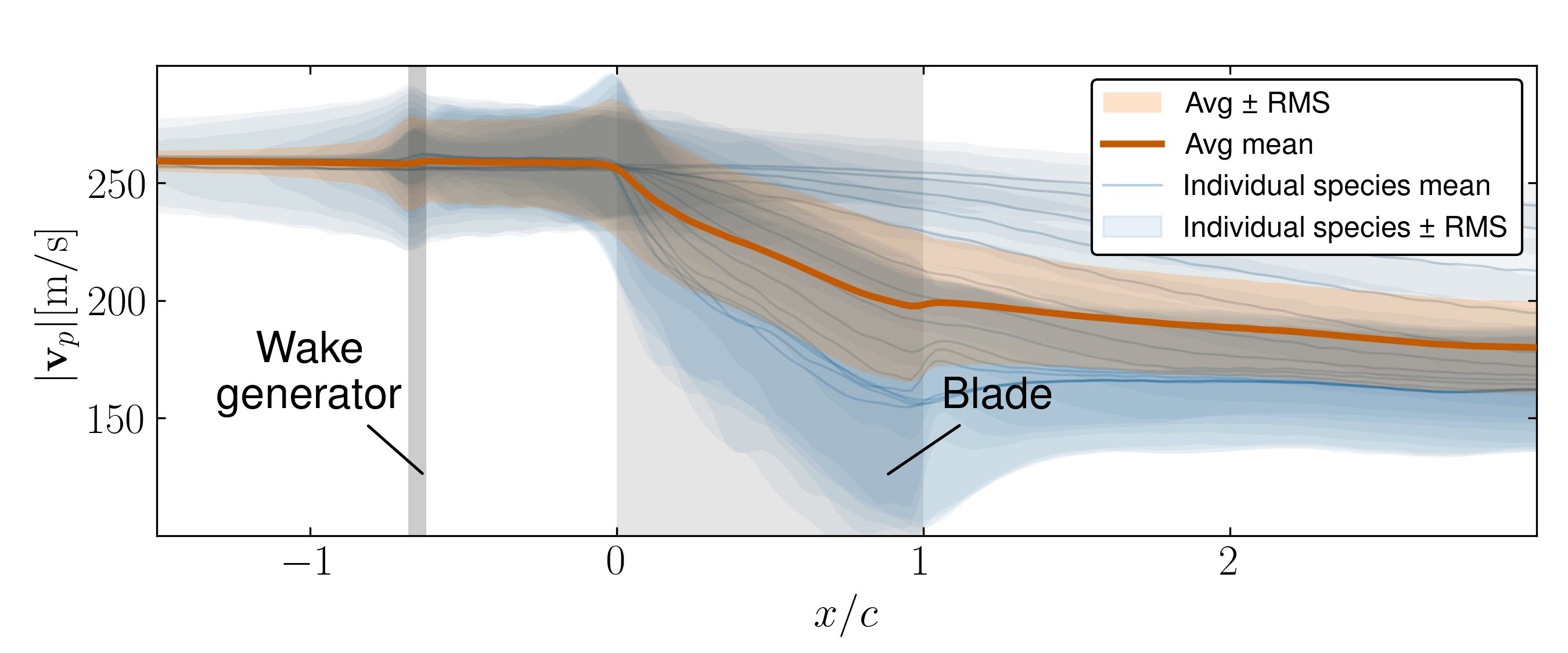}%
  \caption{Mean and standard deviation $\sigma$, indicated by the root mean square (RMS), of the particle velocity against axial position normalized by the chord.}%
  \label{fig:application:sliding:velocity}
\end{figure}

Mean particle velocity $v_p$ and the standard deviation $\sigma$ of the discrete phase as a function of axial position are shown in \cref{fig:application:sliding:velocity}.
Here, darker colors correspond to particles with higher Stokes number, that is stronger inertial effects.
Distinctively visible on the left side of the plot are particles impacting on the wake generator and being reflected upstream, where the associated deceleration and acceleration depend on the local impact angle on the cylinder surface.
A similar effect is observed directly upstream of the blade leading edge.
Particles with lower inertia experience stronger deceleration within the blade passage, with the lightest species remaining in the recirculation region near the trailing edge, identifiable by average particle velocities approaching zero.
Downstream of the blade row, only particles leaving the recirculation region are represented with all considered particle sizes gradually relax toward the mean fluid velocity.
Subsequently, the velocity fluctuations diminish as the wake vortices dissipate into internal energy.

\begin{figure}
  \includegraphics[width=\columnwidth]{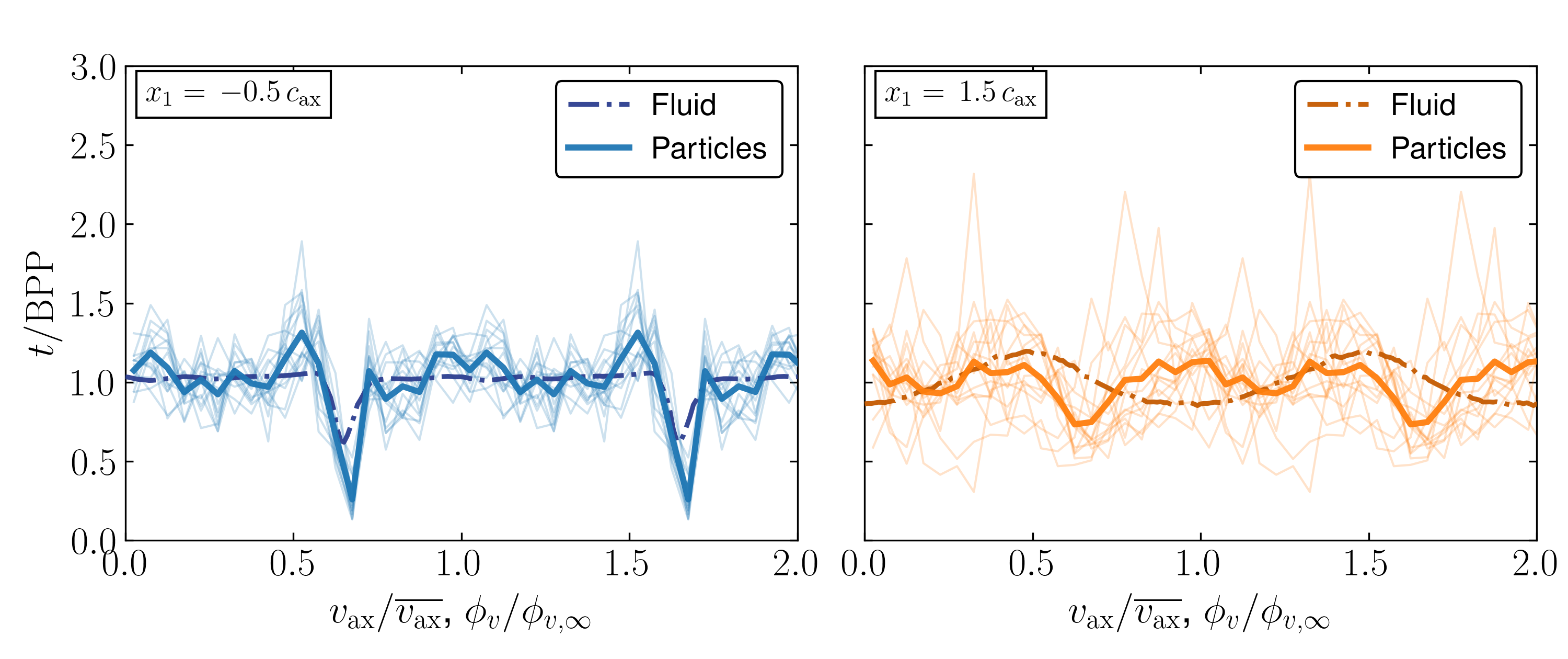}%
  \caption{Fluid velocity and particle flow rate upstream and downstream of the blade row. Individual particle sizes are shaded lightly and the average of all particles in the simulation is highlighted in bold.}%
  \label{fig:application:sliding:wake}
\end{figure}

Despite the disparities between the simulated particle sizes, common features can still be identified in the fluid and particle
flow-rate fields.
Fluid velocity and particle flow rate at distances of \num{0.5} axial chord lengths upstream and downstream of the blade row are illustrated in \cref{fig:application:sliding:wake}.
Here, the upstream position corresponds to a sampling plane slightly downstream of the cylindrical wake generator.
Upstream of the blade row, both the fluid velocity deficit and the minimum of the particle density field appear approximately in phase, with a slight offset of the discrete-phase minimum caused by the motion of the wake generator.
The spanwise extent of the flow rate deficit is narrow and comparable for both phases.
Owing to particle inertia, the discrete-phase wake retains its narrow spatial extent even after traversing the blade passage, whereas the time-averaged fluid wake appears substantially more diffuse.
Despite the differing particle distributions observed in \cref{fig:application:sliding:velocity}, all simulated particle sizes relax towards a comparable wake state already at a distance of just \num{0.5} axial chord lengths downstream of the trailing edge.
The average phase angle between the minima of the fluid and particle fields is approximately \SI{132}{\degree}. Whether this behavior is transferable to different flow configurations or operating conditions remains an open question.

In summary, the presented framework demonstrated its ability to perform scale-resolving simulation and analysis of particle-laden flow even with large displacements.
The relative motion of the wake generations is correctly captured for both the continuous and the discrete phase. By successfully quantifying the phase lag between the both phases, this framework underscores the critical importance of accurately accounting for the temporal variance imposed by upstream stages. These transient effects are heavily influential but are frequently neglected when prescribing simplified inflow conditions in conventional turbomachinery stage simulations.

%% file: 8_conclusion.tex
\section{Conclusion}%
\label{sec:conclusion}

Compressible flow with suspended solid particles in moving domains is present in a wide range of technical applications.
This work presents a high-fidelity Euler--Lagrange framework capable of resolving compressible particle-laden flows within complex, moving geometries.
By coupling a high-order discontinuous Galerkin method for the continuous phase with a discrete Lagrangian point-particle tracking
scheme, the framework achieves high-order accuracy without the prohibitive computational overhead of fully particle-resolved strategies.
To handle moving domains, two prominent moving mesh techniques were successfully integrated: the arbitrary Lagrangian–Eulerian
method for continuous, time-resolved particle-induced wall deformations, and its special case, the sliding mesh approach for rigid
translational or rotational movements along non-conforming interfaces.
Furthermore, the compatibility of these algorithms with unstructured, curved high-order grids enables the seamless handling of complex geometric configurations.
Special emphasis was placed on achieving high-order temporal and spatial accuracy for Lagrangian particle tracking on moving meshes,
specifically when transitioning across non-conforming sliding mesh interfaces.
The capability of the framework was highlighted through its application to two highly challenging compressor rotor blade configurations: the first focusing on complex solid-particle erosion, and the second featuring an upstream cylindrical wake generator.
Although a unified simulation simultaneously leveraging both methodologies is fully supported by the underlying software architecture, these capabilities were evaluated independently here to cleanly isolate their respective performance characteristics, with a combined application left outside the scope of the present study.